\documentclass[aps,prl,numerical,superscriptaddress,showpacs,floatfix,reprint]{revtex4-1}
\usepackage{amsmath,amssymb,mathrsfs,bm} 
\usepackage{url}
\usepackage{hyperref}
\hypersetup{
colorlinks=true,
linkcolor=blue,
anchorcolor =red,
citecolor=blue,
filecolor = red,
urlcolor=blue,
pdfauthor=author}
\usepackage{txfonts}
\usepackage[utf8]{inputenc}
\usepackage{amsmath}
\usepackage{graphicx}
\def\av<#1>{\left\langle\,#1\,\right\rangle}
\def\ev<#1>{\left\langle\,#1\,\right\rangle_{\rm{ev}}}

\begin{document}


\title{Species-Resolved Scaling of Azimuthal Anisotropy: Constraining Attenuation, Collective Expansion, and Hadronic Dynamics in Hydrodynamic Simulations}
\author{ Roy~A.~Lacey}
\email[E-mail: ]{Roy.Lacey@Stonybrook.edu}
\affiliation{Department of Chemistry, 
Stony Brook University, \\
Stony Brook, NY, 11794-3400, USA}
%
%

\date{\today}
\begin{abstract}

Species-resolved azimuthal anisotropy scaling functions are constructed from identified particle $v_2$ and $v_3$ obtained from event-by-event iEBE-VISHNU simulations for Pb+Pb collisions at $\sqrt{s_{NN}}=2.76$ and $5.02$~TeV. The scaling functions exhibit a robust collapse across transverse momentum, centrality, particle species, and beam energy, indicating a common and tightly constrained scaling structure. High scaling fidelity yields quantitative agreement with the data-defined reference through an energy-dependent attenuation baseline $\beta_0$ in central to mid-central collisions and a centrality-dependent modification of the effective attenuation in more peripheral collisions, with only a weak dependence on $\sqrt{s_{NN}}$. The multiplicity dependence of the extracted scaling parameters reflects the interplay of EOS-driven collective expansion, finite system lifetime, and hadronic re-scattering. These results demonstrate that the scaling framework provides a quantitative, constraint-driven probe of the hydrodynamic response, enabling the disentanglement and constraint of the coupled contributions to azimuthal anisotropy.

\end{abstract}

%

\maketitle


Measurements of azimuthal anisotropy in relativistic heavy-ion collisions provide key insights into the properties of the quark-gluon plasma (QGP). These anisotropies are quantified by the Fourier coefficients $v_n$ of the particle azimuthal distribution, defined through
\begin{equation}
V_n \equiv v_n e^{in\Psi_n} = \langle e^{in\phi} \rangle,
\end{equation}
where $v_n$ denotes the magnitude of the $n^{\rm th}$-order anisotropy and $\Psi_n$ is the corresponding event-plane angle~\cite{Bilandzic:2010jr,Luzum:2011mm,Teaney:2012ke}. In practice, the coefficients are measured differentially as $v_n(p_T,\mathrm{cent})$, as a function of transverse momentum $p_T$ and collision centrality, where centrality characterizes the event geometry and system size, typically quantified via charged-particle multiplicity. The coefficients $v_2$ and $v_3$, commonly referred to as elliptic and triangular flow, are particularly sensitive to the initial-state geometry and its fluctuations, as well as to the subsequent dynamical evolution of the medium.

In the low- to intermediate-$p_T$ region, anisotropic flow reflects the hydrodynamic response of the medium to the initial spatial anisotropy, with the transverse-momentum and centrality dependence of $v_n$ governed by the combined influence of viscous attenuation and EOS-driven collective expansion~\cite{Adams:2005dq,Adcox:2004mh,Abelev:2014pua}. At higher transverse momenta, path-length-dependent energy loss (jet quenching) also contributes to the observed anisotropy, further linking $v_n$ to the transport properties of the medium.

Hydrodynamic models, particularly those coupled to hadronic transport, have achieved considerable success in describing the measured $v_n(p_T,\mathrm{cent})$ for identified particles across a range of collision systems and energies. In practice, such hydrodynamic calculations are most commonly applied to the low- to intermediate-$p_T$ region where collective dynamics dominate the anisotropic response. Representative examples include hybrid frameworks such as VISHNU and iEBE-VISHNU, as well as other implementations combining viscous hydrodynamics with hadronic afterburners (e.g., MUSIC coupled to UrQMD or SMASH)~\cite{Song:2010aq,Shen:2015qta,Zhao:2017yhj,Schenke:2010nt,Schenke:2011bn,McDonald:2016vlt,Bleicher:1999xi,Bass:1998ca,SMASH:2016zqf}. These models incorporate the essential ingredients of the evolution, including initial-state geometry, QGP transport properties such as $\eta/s$, the equation of state, and late-stage hadronic re-scattering. However, the resulting $v_n(p_T,\mathrm{cent})$ reflect a convolution of these effects, making it challenging to isolate their relative contributions from the calculated distributions alone.

In this context, scaling approaches have emerged as a powerful tool for revealing the underlying response structure of anisotropic flow in the low- to intermediate-$p_T$ region. In particular, species-resolved scaling relations show that the transverse-momentum and centrality dependence of $v_n$ for different particle species collapses onto common scaling functions~\cite{Lacey:2024fpb,Lacey:2024uky}. This behavior indicates that the observed anisotropy is governed by a limited set of parameters encoding the effects of viscous attenuation, EOS-driven collective expansion, and particle-dependent interactions. The framework constrains the structure of the hydrodynamic response through simultaneous requirements on species-resolved scaling, inter-harmonic consistency, and system-size dependence, rather than relying on a unique determination of individual parameters. While complementary observables, including fluctuation-based measures of radial flow and transverse-momentum correlations, provide important constraints~\cite{ALICE:2025iud,ATLAS:2025ztg,Du:2025dpu}, their sensitivity to overlapping dynamical effects limits their ability to uniquely disentangle the coupled contributions of viscous attenuation, collective expansion, and hadronic re-scattering.

Jet quenching can also contribute to anisotropy at higher transverse momenta through path-length-dependent energy loss; however, the present analysis emphasizes the low- to intermediate-$p_T$ region. The scaling framework accommodates the full $p_T$ range and provides a unified description linking collective flow and high-$p_T$ behavior~\cite{Lacey:2024fpb,Lacey:2024uky}. Agreement with these scaling functions provides a more differential constraint than the reproduction of $v_n(p_T,\mathrm{cent})$ alone, enabling an assessment of whether the underlying balance of collective expansion, dissipation, and hadronic dynamics is realized consistently within a given model.

Within this framework, a complementary approach to hydrodynamic modeling is achieved through a differential characterization of the underlying dynamics. By expressing the anisotropic flow coefficients in terms of reduced response functions and scaling variables, the contributions from viscous attenuation, EOS-driven collective expansion, and hadronic re-scattering are systematically isolated through distinct scaling parameters. This formulation establishes a quantitative mapping between the observed scaling behavior and the underlying dynamical contributions to collective flow.

In this work, the species-resolved scaling framework is applied to the anisotropic flow coefficients $v_2$ and $v_3$ obtained from event-by-event iEBE-VISHNU hydrodynamic simulations~\cite{Zhao:2017yhj} for identified pions, kaons, and protons in Pb+Pb collisions at $\sqrt{s_{NN}}=2.76$ and $5.02$~TeV \cite{Zhu:2019twz,ALICE:2022zks,ALICE:2014wao,ALICE:2016cti,ALICE:2017nuf,ALICE:2018lao}. By comparing the scaling functions constructed from the hydrodynamic calculations with those inferred from experimental data~\cite{ALICE:2011ab,Lacey:2024uky}, the analysis tests the robustness of the scaling structure and constrains how the interplay of QGP transport, EOS-driven collective expansion, and hadronic re-scattering is realized across centrality. In this way, the scaling framework serves both as a validation of the hydrodynamic description and as a diagnostic tool for how the underlying mechanisms governing anisotropic flow are realized.


The species-resolved scaling framework follows the formulation developed in Refs.~\cite{Lacey:2024fpb,Lacey:2024uky}, anchored to the kaon scaling function in ultra-central (uc, 0--1\%) Pb+Pb collisions at $\sqrt{s_{NN}}=5.02$~TeV, which serves as a universal reference. This choice preserves species identity and enables direct comparison of mass-, baryon-number–, and coupling-dependent responses within a common medium.

In this framework, anisotropic flow is expressed in terms of the reduced response $v_n/\varepsilon_n$, with geometric inputs specified by the eccentricities $\varepsilon_n(\mathrm{cent})$ and the transverse size scale $\mathbb{R} \propto \langle N_{\rm ch} \rangle^{1/3}$. The effective scaling response is governed by $\beta = k_\beta \beta_0$, where $\beta_0$ denotes the reference attenuation and $k_\beta$ quantifies system- and centrality-dependent modifications. Viscous corrections are incorporated through $\delta f = \kappa p_T^2$~\cite{Liu:2018hjh}, leading to a characteristic dependence on $(n + \kappa p_T^2)$ and motivating the use of the transverse kinetic energy ${\rm KE_T} \equiv m_T - m_0$ as a scaling variable. The resulting representation reduces mass-dependent kinematic effects and organizes the attenuation into an approximately linear dependence in the flow-dominated regime when expressed in the scaling representation.

Species dependence enters through two coefficients associated with late-stage dynamics. Meson re-scattering is quantified by $\zeta_{\rm hs}$, defined relative to the kaon baseline, with $\zeta_m = 1 - \zeta_{\rm hs}$. The baryon response to collective expansion is governed by $\zeta_{\rm rf}$, which encodes the species-dependent influence of radial flow, with $\zeta_b = (1 - \zeta_{\rm rf})^{|n_B|}$, where $n_B$ denotes the baryon number. Quantities labeled with primes refer to the comparison system or centrality, while unprimed quantities correspond to the reference system.

For mesons, the $v_2$ scaling relation is
\begin{multline}
\frac{v_2(p_T,\mathrm{uc})}{\varepsilon_2(\mathrm{uc})}\,
e^{\tfrac{2\beta_0}{\mathbb{R}_{\rm uc}}(2+\kappa p_T^2)}
=
e^{\alpha\,\tfrac{2\beta_0}{\mathbb{R}_{\rm uc}}
(2+\kappa p_T^2)\,\zeta_M^{(2)}}
\\[-2pt]
\times
\left(\frac{v_2'(p_T)}{\varepsilon_2'}\right)^{\zeta_m}\,
e^{\tfrac{2\zeta_m\beta}{\mathbb{R}_{\rm uc}}
\left(\tfrac{\mathbb{R}_{\rm uc}}{\mathbb{R}'}-1\right)
(2+\kappa p_T^2)},
\label{eq:v2_scaling_mesons}
\end{multline}
where $\alpha$ controls the contribution of the reference-side exponential prefactor. For ultra-central collisions, $\alpha=1$ and the full $p_T$-dependent prefactor is retained. For non-ultra-central selections, this prefactor contributes only to an overall normalization, while near-uc (0--5\%) selections interpolate smoothly between these limits. The normalization exponent is
$\zeta_M^{(2)}=\zeta_m+\gamma_{32}^{X}+(1-k_\beta)$,
where $\gamma_{32}^{X} \equiv \ln\!\left[(\varepsilon_3/\varepsilon_2)_{\mathrm{ref}}/(\varepsilon_3/\varepsilon_2)_{X}\right]$ is a geometry-only offset quantifying differences in the eccentricity ratio $\varepsilon_3/\varepsilon_2$ between the reference and comparison systems within the same centrality class. For the ultra-central reference system and the beam energies considered here, $\gamma_{32}^{X}\approx 0$, reflecting the near invariance of the eccentricity ratio across systems in this limit.

An additional constraint is provided by the inter-harmonic mapping
\begin{multline}
\frac{v_2(p_T,\mathrm{uc})}{\varepsilon_2(\mathrm{uc})}\,
e^{\tfrac{2\beta_0}{\mathbb{R}_{\rm uc}}(2+\kappa p_T^2)}
=
e^{\tfrac{2\beta_0}{\mathbb{R}_{\rm uc}}\,
\zeta_M^{(2\leftarrow3)}}
\\[-2pt]
\times
\left(\frac{v_3'(p_T)}{\varepsilon_3'}\right)^{\tfrac{2}{3}\zeta_m}\,
e^{\tfrac{2\zeta_m\beta}{\mathbb{R}_{\rm uc}}
\left(\tfrac{\mathbb{R}_{\rm uc}}{\mathbb{R}'}
-1\right)(3+\kappa p_T^2)},
\label{eq:v2_from_v3_mesons}
\end{multline}
with $\zeta_M^{(2\leftarrow3)}=1-\zeta_M^{(2)}$. The inter-harmonic mapping retains the same attenuation and size-mismatch structure as the intra-harmonic scaling relation, ensuring that both constraints are governed by a common attenuation and medium-response structure.

For baryons, the corresponding $v_2$ scaling relation is
\begin{multline}
\frac{v_2(p_T,\mathrm{uc})}{\varepsilon_2(\mathrm{uc})}\,
e^{\tfrac{2\beta_0}{\mathbb{R}_{\rm uc}}(2+\kappa p_T^2)}
=
e^{(1-\alpha)\,\tfrac{2\beta_0}{\mathbb{R}_{\rm uc}}
(2+\kappa p_T^2)\,\zeta_B^{(2)}}
\\[-2pt]
\times
\left(\frac{v_2'(p_T)}{\varepsilon_2'}\right)^{\zeta_b}\,
e^{\tfrac{2\zeta_b\beta}{\mathbb{R}_{\rm uc}}
\left(\tfrac{\mathbb{R}_{\rm uc}}{\mathbb{R}'}
-1\right)(2+\kappa p_T^2)},
\label{eq:v2_scaling_baryons}
\end{multline}
where $\zeta_B^{(2)}=-\zeta_b(|n_B|/k_\beta-\gamma_{32}^{X})$.

An additional constraint is provided by the inter-harmonic mapping
\begin{multline}
\frac{v_2(p_T,\mathrm{uc})}{\varepsilon_2(\mathrm{uc})}\,
e^{\tfrac{2\beta_0}{\mathbb{R}_{\rm uc}}(2+\kappa p_T^2)}
=
e^{\tfrac{2\beta_0}{\mathbb{R}_{\rm uc}}\,
\zeta_B^{(2\leftarrow3)}}
\\[-2pt]
\times
\left(\frac{v_3'(p_T)}{\varepsilon_3'}\right)^{\tfrac{2}{3}\zeta_b}\,
e^{\tfrac{2\zeta_b\beta}{\mathbb{R}_{\rm uc}}
\left(\tfrac{\mathbb{R}_{\rm uc}}{\mathbb{R}'}
-1\right)(3+\kappa p_T^2)},
\label{eq:v2_from_v3_baryons}
\end{multline}
with $\zeta_B^{(2\leftarrow3)}=-\zeta_b(2|n_B|/k_\beta+\gamma_{32}^{X})$.

Together, these relations define a closed, species-resolved constraint system linking attenuation, geometry, and late-stage dynamics. The common attenuation baseline and explicit size-mismatch structure govern the transverse-momentum dependence, while species-dependent normalization terms encode the influence of radial flow and hadronic re-scattering through $\zeta_{\rm rf}$ and $\zeta_{\rm hs}$ on the effective scaling response. In this way, the framework separates collective expansion and dissipative effects, enabling a differential and quantitative characterization of the underlying medium dynamics.


The present study employs results from event-by-event viscous hydrodynamic simulations performed within the VISHNU hybrid framework~\cite{Song:2010aq,Shen:2015qta}. The anisotropy coefficients used in this analysis correspond to identified pions, kaons, and protons in Pb+Pb collisions at $\sqrt{s_{NN}}=2.76$ and $5.02$~TeV, obtained from iEBE-VISHNU simulations with AMPT initial conditions and compared to ALICE measurements~\cite{Zhao:2017yhj}. This framework combines relativistic viscous hydrodynamics for the QGP phase with a microscopic hadronic transport model (UrQMD)~\cite{Bleicher:1999xi,Bass:1998ca} for the late-stage evolution, providing a unified dynamical description of partonic and hadronic evolution.

In this approach, the key ingredients governing anisotropic flow are encoded through the initial-state geometry, medium properties, and late-stage dynamics. The geometry is characterized by fluctuating eccentricities $\varepsilon_n$, while the system size and energy density are determined by the initial entropy density profile constrained by the measured charged-particle multiplicity. The hydrodynamic evolution of the QGP phase is governed by the specific shear viscosity $\eta/s$ and the equation of state (EOS), which together control viscous attenuation and EOS-driven collective expansion. The subsequent transition to the hadronic phase at a switching temperature $T_{\rm sw} \sim 148$~MeV is described by microscopic transport dynamics (UrQMD), including hadronic re-scattering, resonance formation, and decay.

While hadronic re-scattering does not modify the QGP shear viscosity $\eta/s$, it contributes additional dissipation to the final-state anisotropy. Consequently, the effective scaling response reflects the combined influence of QGP viscosity, EOS-driven expansion, the system lifetime governing their duration, and late-stage hadronic dynamics. Within the scaling framework, these contributions are encoded through the parameters $\beta_0$ and $k_\beta$, which characterize the effective attenuation, together with $\zeta_{\rm rf}$ and $\zeta_{\rm hs}$, which quantify the contributions from radial flow and hadronic re-scattering, respectively. These parameters represent effective, system-dependent quantities that reflect the evolving balance of collective expansion, dissipation, and hadronic dynamics, and are therefore expected to depend on centrality, multiplicity, and beam energy.

This provides a controlled setting in which the underlying dynamical ingredients are known, enabling a direct test of whether the species-resolved scaling relations recover the structure of the hydrodynamic response. In particular, the reduced-response scaling probes how attenuation and geometric effects combine through the parameter $\beta$, while the species-dependent normalization encodes late-stage dynamics. The reference attenuation scale $\beta_0$, fixed by the data-defined baseline, provides a common normalization against which the effective scaling response from the hydrodynamic calculations can be assessed, enabling a direct test of whether the scaling framework can disentangle the underlying dynamical contributions to the hydrodynamic response, thereby placing mechanistic constraints on the interpretation of the data-derived scaling functions.
\begin{figure*}[tbh]
    \centering
    \includegraphics[clip,width=0.75\linewidth]{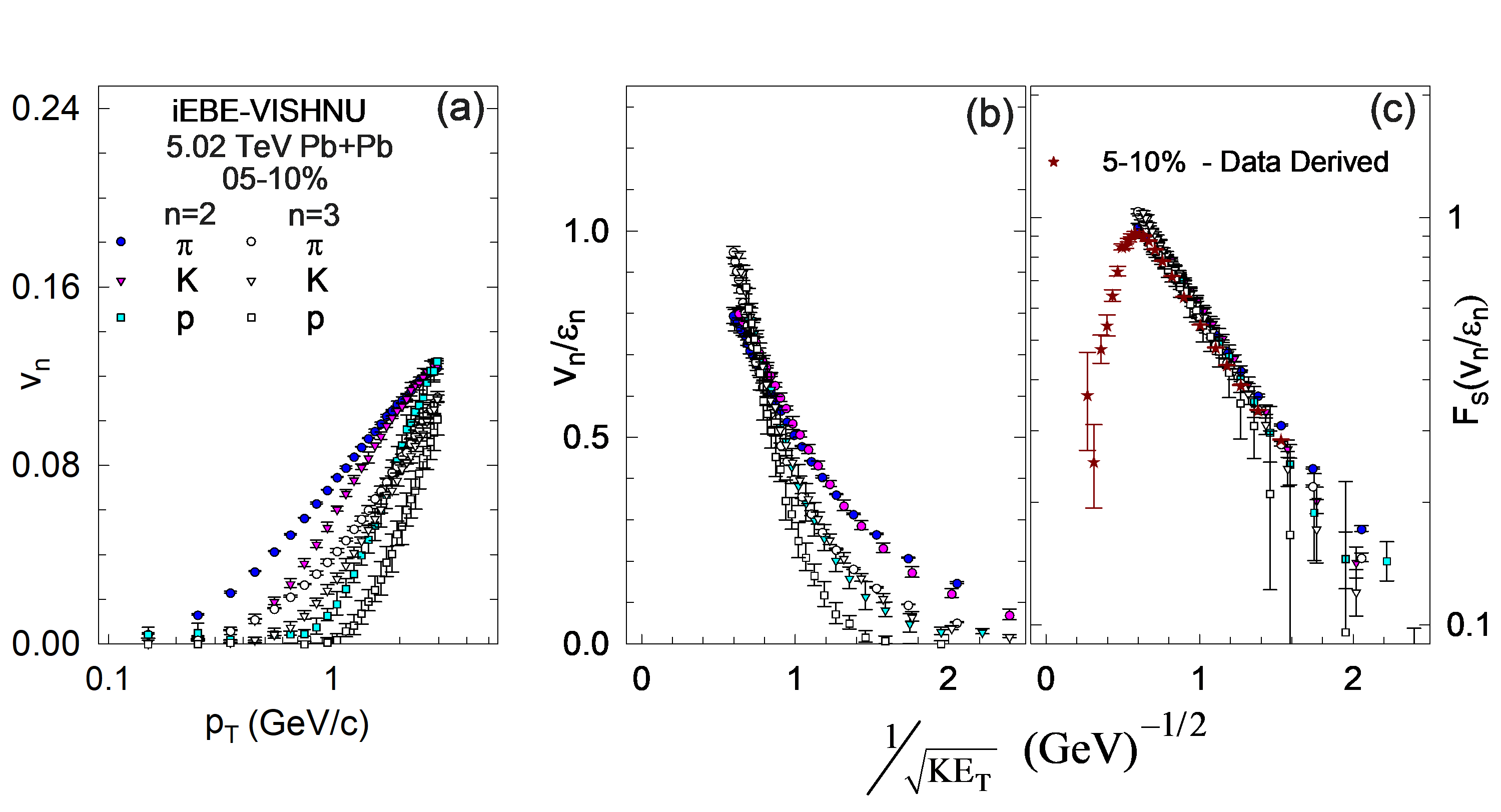} \\
		\includegraphics[clip,width=0.75\linewidth]{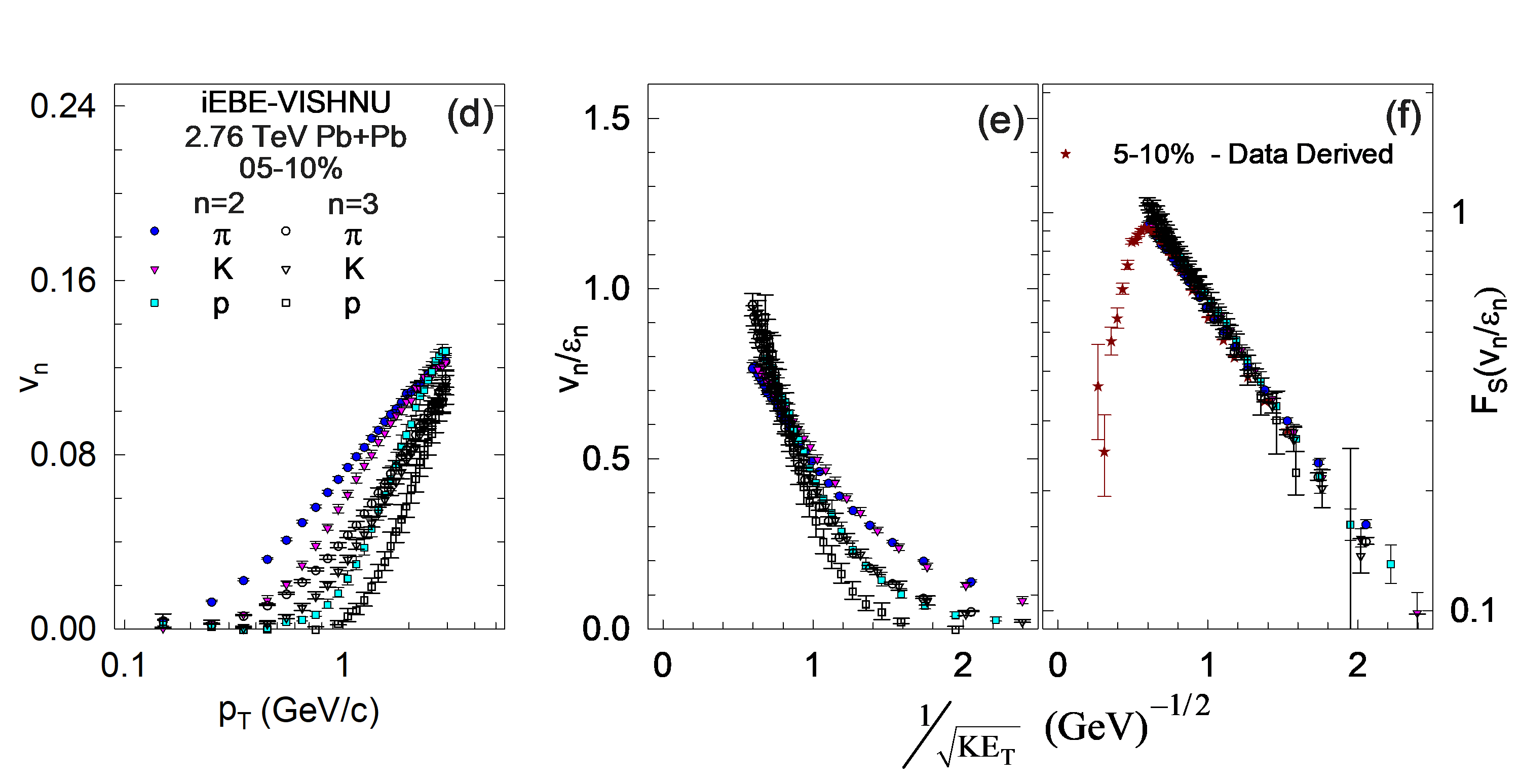}
\caption{Species-resolved scaling for identified pions, kaons, and protons in Pb+Pb collisions at $\sqrt{s_{NN}}=5.02$~TeV [(a)--(c)] and $2.76$~TeV [(d)--(f)]. Panels (a,d) show the anisotropy coefficients $v_{2,3}(p_T)$ from event-by-event iEBE-VISHNU simulations~\cite{Zhao:2017yhj}, while panels (b,e) show the reduced response $v_{2,3}/\varepsilon_{2,3}$. Panels (c,f) display the scaling functions $F_s(v_n/\varepsilon_n)$ versus $1/\sqrt{{\rm KE_T}}$ (${\rm KE_T} \equiv m_T - m_0$). Filled stars denote scaling functions determined from experimental measurements~\cite{Lacey:2024uky}.
}
    \label{fig:1}
\end{figure*}
\begin{figure*}[tbh]
    \centering
    \includegraphics[clip,width=0.75\linewidth]{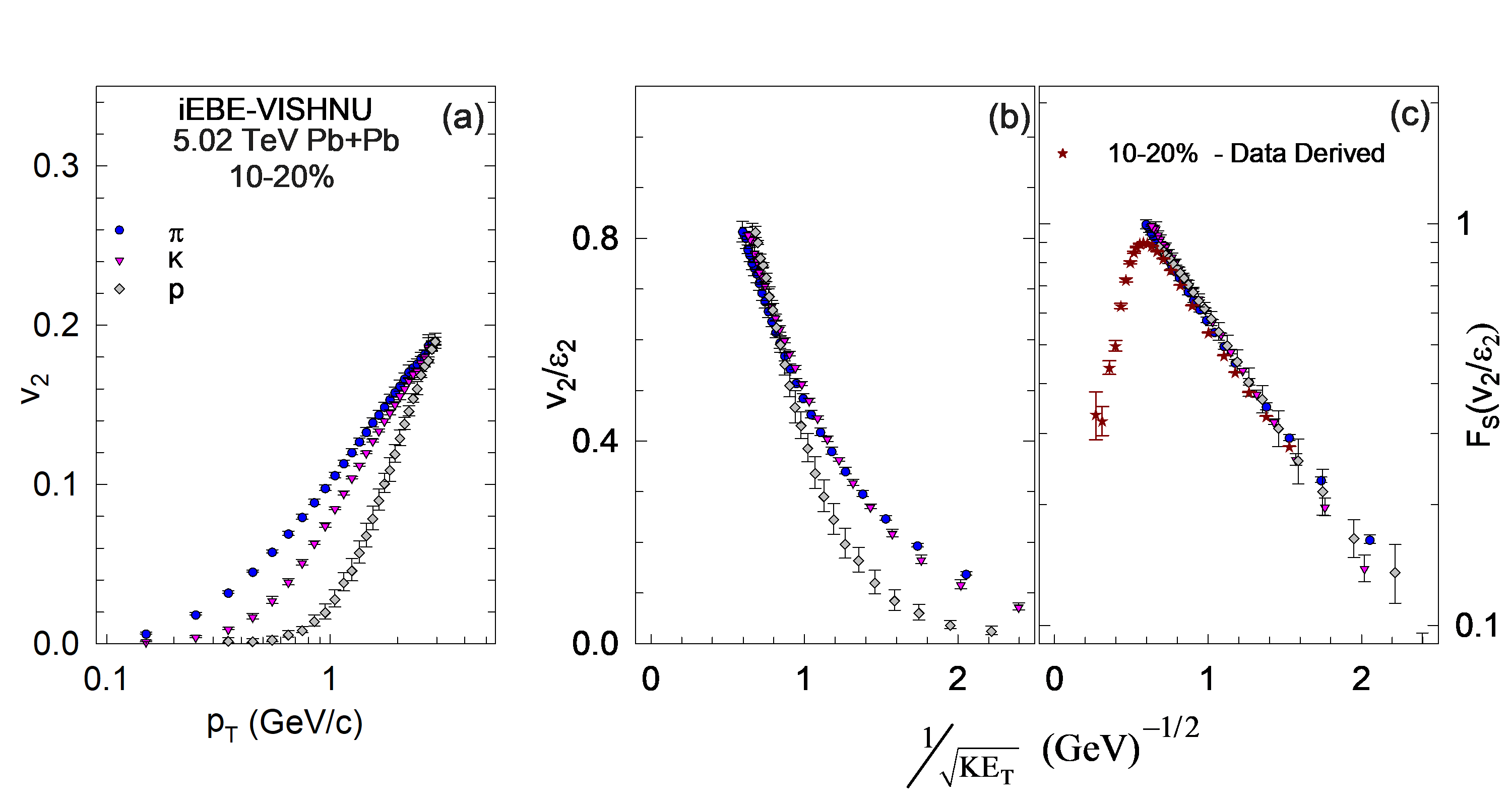} \\
		\includegraphics[clip,width=0.75\linewidth]{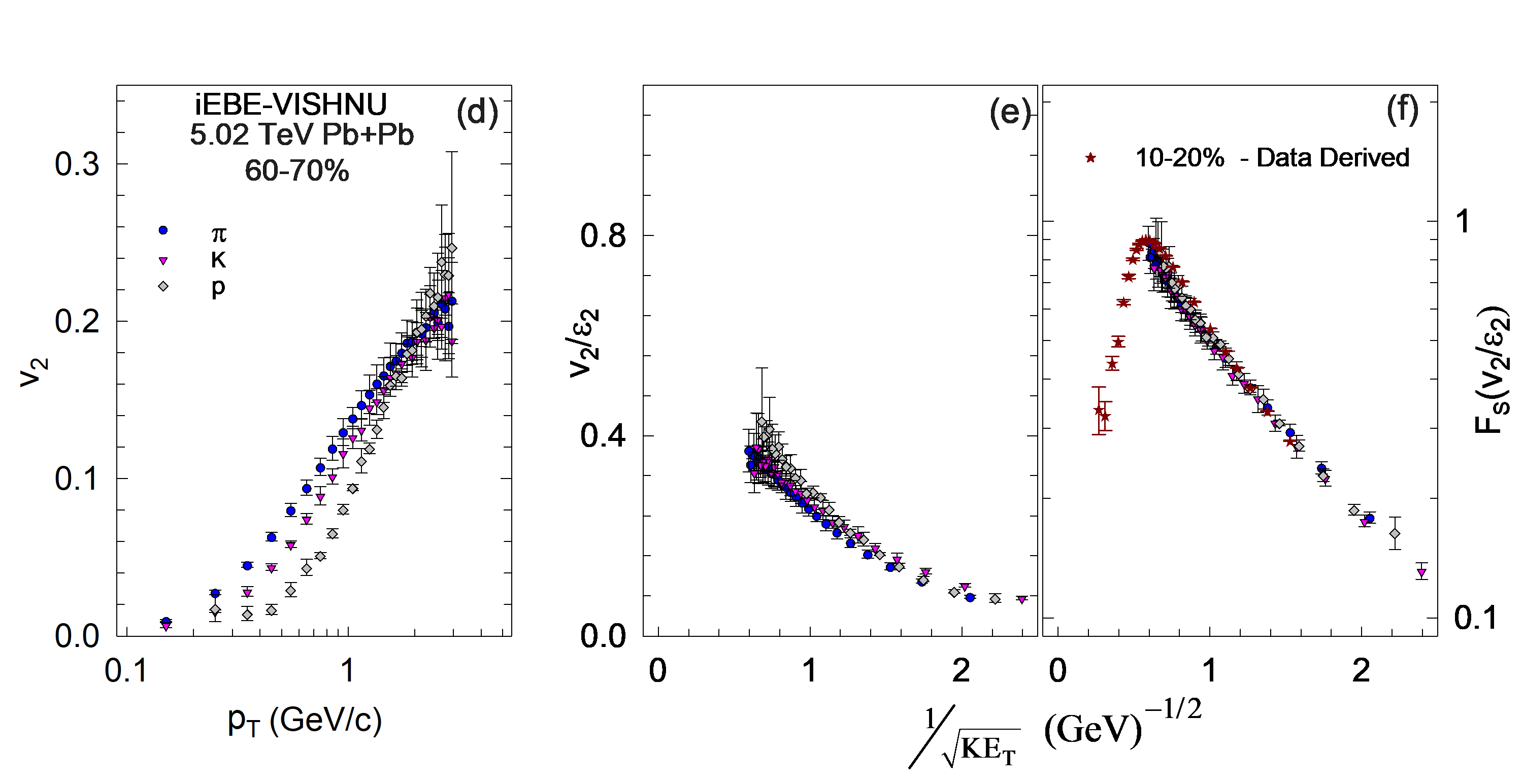}
\caption{$v_2$ scaling for identified pions, kaons, and protons in Pb+Pb collisions at $\sqrt{s_{NN}}=5.02$~TeV. Panels (a)--(c) correspond to 10--20\% central collisions, while panels (d)--(f) correspond to 60--70\% collisions. Panels (a,d) show $v_2(p_T)$, panels (b,e) show the reduced response $v_2/\varepsilon_2$, and panels (c,f) display the scaling functions $F_s(v_2/\varepsilon_2)$ versus $1/\sqrt{{\rm KE_T}}$ (${\rm KE_T}\equiv m_T - m_0$). Filled stars denote scaling functions determined from experimental measurements~\cite{Lacey:2024uky}.
}
    \label{fig:2}
\end{figure*}
\begin{figure*}[tbh]
    \centering
    \includegraphics[clip,width=0.75\linewidth]{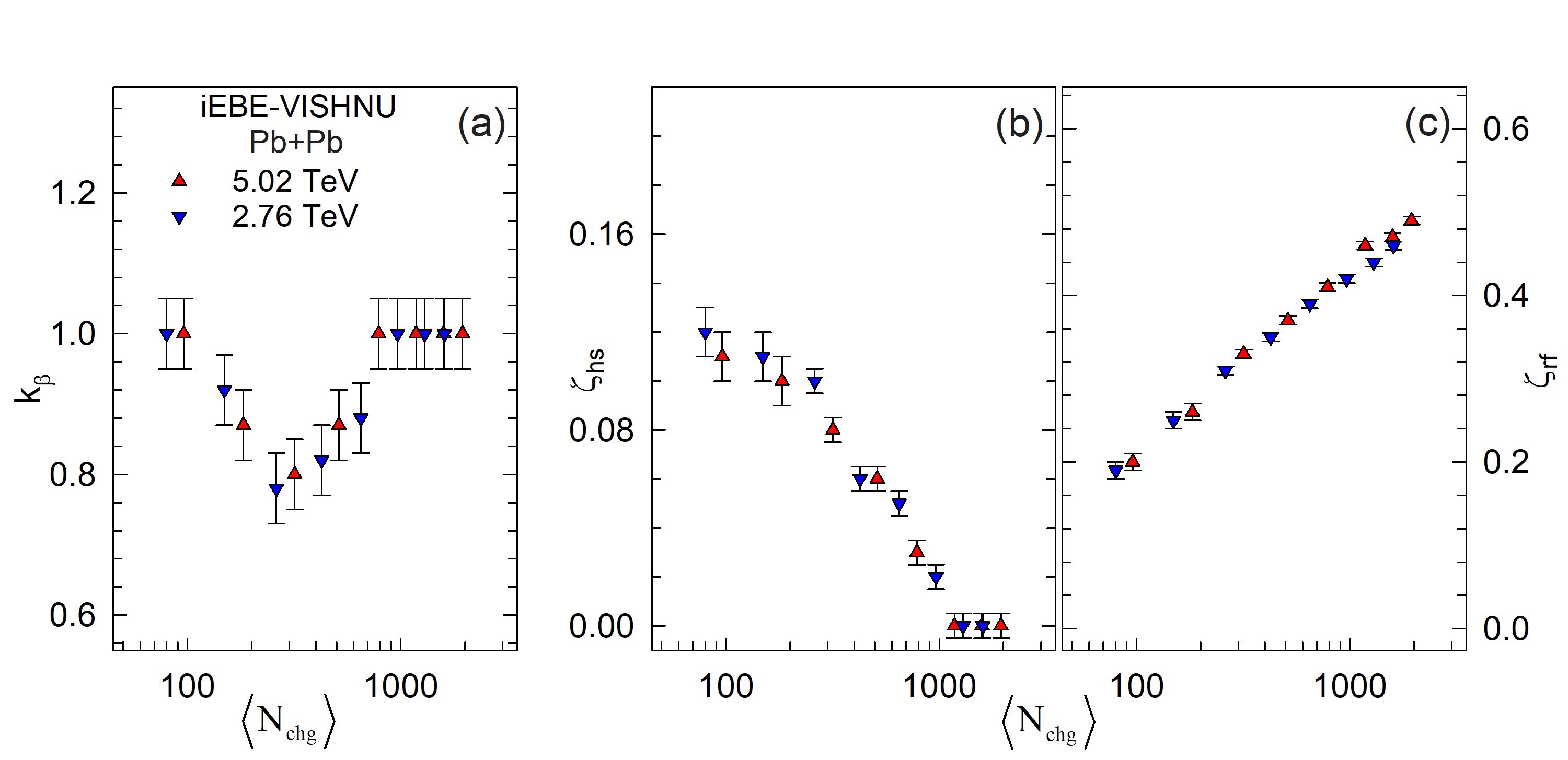}
\caption{Multiplicity dependence of the extracted scaling parameters for Pb+Pb collisions at $\sqrt{s_{NN}}=2.76$ and $5.02$~TeV. Panel (a) shows the attenuation parameter $k_\beta$, panel (b) the hadronic re-scattering parameter $\zeta_{\rm hs}$, and panel (c) the radial flow parameter $\zeta_{\rm rf}$, each plotted as a function of the mean charged-particle multiplicity $\langle N_{\rm chg} \rangle$, which serves as a proxy for system size and energy density.}
    \label{fig:3}
\end{figure*}
%


The species-resolved scaling framework is applied to event-by-event hydrodynamic simulations to quantitatively constrain the medium response encoded in the anisotropy coefficients $v_2(p_T,\mathrm{cent})$ and $v_3(p_T,\mathrm{cent})$. The anisotropy coefficients $v_2(p_T,\mathrm{cent})$ and $v_3(p_T,\mathrm{cent})$ for identified pions, kaons, and protons from iEBE-VISHNU simulations~\cite{Zhao:2017yhj} are shown in Figs.~\ref{fig:1}(a)--(f) for representative central (5--10\%) Pb+Pb collisions at $\sqrt{s_{NN}}=5.02$~TeV [(a)--(c)] and $2.76$~TeV [(d)--(f)]. These calculations provide a good description of the measured $v_{2,3}(p_T)$ in this centrality range. A clear mass ordering is observed at low transverse momentum—most prominently for $v_2$, reflecting its sensitivity to collective expansion—while $v_3$, which is subject to stronger viscous attenuation, exhibits a weaker but still discernible species dependence. At intermediate $p_T$, the behavior of $v_{2,3}(p_T)$ reflects the interplay between radial flow and viscous attenuation, leading to a progressive reduction of the mass ordering. While this agreement indicates that the hydrodynamic calculations capture the overall magnitude of the anisotropic flow, it does not uniquely constrain the underlying attenuation, as similar $v_n(p_T)$ can arise from different combinations of viscous attenuation and collective expansion.

To isolate the medium-driven response, the reduced response $v_{2,3}/\varepsilon_{2,3}$ for the same centrality selection, shown in Figs.~\ref{fig:1}(b) and (e), removes the dominant geometric contribution associated with the initial-state eccentricities. In this representation, the species-dependent separation—arising from the combined influence of radial flow and viscous attenuation—becomes more transparent. However, the absence of a universal scaling across particle species and harmonics indicates that geometry alone is insufficient to account for the observed behavior. This motivates the application of the full scaling framework to disentangle the respective roles of attenuation, radial flow, and hadronic re-scattering in shaping the effective scaling response, providing sensitivity to the underlying response that is not accessible through direct comparisons of $v_n(p_T,\mathrm{cent})$ alone.

The full scaling relations [Eqs.~(\ref{eq:v2_scaling_mesons})--(\ref{eq:v2_from_v3_baryons})] yield the scaling functions ${\rm F_s}(v_n/\varepsilon_n)$ shown in Figs.~\ref{fig:1}(c) and (f), where the hydrodynamic results are compared to the data-defined reference scaling functions. For both collision energies, the hydrodynamic results exhibit a clear species-resolved collapse within the adopted scaling representation, demonstrating a common transverse-momentum dependence. The simultaneous scaling of $v_2$ and $v_3$, together with the mapping of $v_3$ onto $v_2$, provides a stringent constraint on the initial-state geometry and the associated medium response, providing a non-trivial and over-constrained test of the underlying medium response. High scaling fidelity is achieved for both beam energies with $k_\beta=1$, with $\beta_0^{\rm hydro} \approx 0.85\,\beta_0$ at 5.02~TeV and $\beta_0^{\rm hydro} \approx 0.80\,\beta_0$ at 2.76~TeV, together with $\zeta_{\rm hs}\approx0$ and values of $\zeta_{\rm rf}$ consistent with those extracted from data at each respective beam energy. The offset between the data-defined and hydrodynamic attenuation baselines reflects a modest difference in the effective time-integrated dissipation, arising from model-dependent choices for transport coefficients, expansion dynamics, and hadronic evolution, rather than a change in the underlying scaling structure. These values indicate that the hydrodynamic calculations realize a near-reference balance of viscous attenuation, collective expansion, and hadronic dynamics in central to mid-central collisions. This consistency indicates that the scaling structure exhibits only a weak dependence on $\sqrt{s_{NN}}$. A modest overshoot of the scaling functions relative to the data-defined reference is observed at high values of ${\rm KE_T}$, consistent with the onset of jet-quenching effects not included in the hydrodynamic calculations. This observation underscores that the scaling framework extends beyond the hydrodynamic regime, providing a unified description across the full transverse-momentum range that connects collective flow to quenching-dominated dynamics. In this context, the comparison demonstrates that the scaling analysis not only constrains the hydrodynamic response, but also provides mechanistic constraints that anchor the physical interpretation of the data-defined scaling functions, demonstrating that the framework constrains how these coupled effects are realized rather than simply reflecting their presence in hydrodynamic modeling.

For the data-defined reference system (ultra-central Pb+Pb at $\sqrt{s_{NN}}=5.02$~TeV), the attenuation baseline is fixed to $\beta_0=0.88$ with $k_\beta=1$ and $\zeta_{\rm hs}\approx0$, reflecting the negligible influence of hadronic re-scattering in this regime. In the hydrodynamic calculations, for central to mid-central collisions (0--5\% to $\sim$20--30\%), high-fidelity scaling is achieved for both beam energies with $k_\beta=1$, with $\beta_0^{\rm hydro}\approx0.85\,\beta_0$ at 5.02~TeV and $\beta_0^{\rm hydro}\approx0.80\,\beta_0$ at 2.76~TeV. Over this centrality range, $\zeta_{\rm hs}\approx0$ at 5.02~TeV and remains small at 2.76~TeV, while $\zeta_{\rm rf}$ is consistent with the corresponding data-extracted values at each energy. These results indicate that the hydrodynamic calculations reproduce the observed anisotropy with a near-reference attenuation baseline and a similar balance of collective expansion and dissipation at both energies, demonstrating only a weak dependence of the effective scaling response on $\sqrt{s_{NN}}$. In more peripheral collisions, deviations emerge, where quantitative agreement requires a centrality-dependent effective attenuation, $\beta^{\rm hydro}=k_\beta\beta_0^{\rm hydro}$, reflecting a modified scaling response relative to the central baseline.

The centrality dependence of this behavior is illustrated by the $v_2$ scaling shown in Fig.~\ref{fig:2}, where panels (a)--(c) correspond to 10--20\% central collisions and panels (d)--(f) to 60--70\% collisions at $\sqrt{s_{NN}}=5.02$~TeV. The $v_2(p_T)$ and reduced-response panels (a,b) exhibit a pronounced separation between baryons and mesons—most clearly seen in the reduced-response panel (b)—reflecting strong radial flow, whereas the corresponding peripheral panels (d,e) show a substantially reduced baryon--meson separation, with the contrast particularly evident in panel (e), consistent with a weaker radial flow response.

By contrast, the scaling function panels (c) and (f), corresponding to $F_s(v_2/\varepsilon_2)$, compare the hydrodynamic results to the data-defined reference scaling functions and exhibit a collapse across particle species for both centrality classes, demonstrating a common transverse-momentum dependence. In the peripheral regime, however, achieving high-fidelity scaling relative to the data-defined reference requires a centrality-dependent effective attenuation, $\beta^{\rm hydro}=k_\beta^{\rm hydro}\,\beta_0^{\rm hydro}$. While the scaling collapse remains robust, the agreement with the data-defined scaling functions becomes less precise in the most peripheral collisions. The extracted values of $k_\beta^{\rm hydro}$ indicate a reduction in the effective attenuation in mid-peripheral collisions (40--50\%, $k_\beta^{\rm hydro}\sim0.80$) relative to both more central and more peripheral collisions, followed by a recovery toward the central baseline at higher peripheralities. This behavior indicates that the underlying scaling structure is preserved, while the effective attenuation reflects a centrality-dependent modification of the accumulated viscous response. These trends are further elucidated by the multiplicity dependence of the extracted scaling parameters, discussed below in Fig.~\ref{fig:3}.

The underlying origin of this centrality dependence is reflected in the behavior of the species-dependent scaling parameters. The hadronic re-scattering parameter $\zeta_{\rm hs}$ increases toward more peripheral collisions, indicating a growing influence of late-stage hadronic dynamics, while the extracted radial flow parameter $\zeta_{\rm rf}$ remains consistent between hydrodynamic calculations and data, indicating that the EOS-driven collective expansion is tightly constrained across centrality. These trends provide a mechanistic interpretation of the centrality-dependent modification of the effective scaling response, wherein the redistribution of attenuation arises from the interplay between reduced system lifetime and enhanced hadronic re-scattering.

This interpretation also provides context for the known limitations of hydrodynamic calculations in more peripheral collisions. While the unscaled hydrodynamic results deviate from the data in this regime, the scaled distributions exhibit a robust collapse, although the agreement with the data-defined scaling functions becomes less precise in the most peripheral collisions. This demonstrates that the underlying scaling structure remains robust, and that the discrepancy primarily reflects how the effective scaling response is realized within the hydrodynamic model rather than a breakdown of the scaling framework itself. This separation between robust scaling structure and the model-dependent realization of the response is a central feature of the framework.

Further insight is provided by the multiplicity dependence of the scaling parameters shown in Fig.~\ref{fig:3}. The extracted values indicate that $\zeta_{\rm hs}$ remains small for central to mid-central collisions and increases toward more peripheral collisions, while $\zeta_{\rm rf}$ increases with multiplicity, consistent with stronger pressure gradients and enhanced collective expansion in high-density events. These trends are consistent across both beam energies, with only modest differences in magnitude.

Taken together, these observations provide a coherent picture of the centrality dependence of the effective scaling response. Strong radial flow in high-multiplicity collisions is associated with near-reference attenuation, while in more peripheral collisions the reduced system lifetime limits the accumulation of viscous attenuation and enhances the relative importance of hadronic re-scattering. The resulting variation of $k_\beta$ reflects a redistribution of the effective response across centrality.

Collectively, these results demonstrate that the species-resolved scaling structure is robust across collision energies and system conditions, and that deviations from the reference scaling function reflect a centrality-dependent modification of the effective attenuation rather than a breakdown of the scaling framework. The observed behavior of $k_\beta$ encodes the interplay of EOS-driven collective expansion, finite system lifetime, and late-stage hadronic dynamics, establishing the scaling framework as a quantitative, constraint-driven probe of the underlying hydrodynamic response. The consistency of these trends across beam energies and centrality classes indicates that the extracted scaling structure reflects generic features of the hydrodynamic response rather than model-specific details.

In summary, the species-resolved scaling framework is applied to azimuthal anisotropy coefficients from event-by-event iEBE-VISHNU hydrodynamic simulations for Pb+Pb collisions at $\sqrt{s_{NN}}=2.76$ and $5.02$~TeV. The results demonstrate a robust collapse of the scaled anisotropy across transverse momentum, centrality, particle species, and beam energy, indicating a common and highly constrained scaling structure. High scaling fidelity yields quantitative agreement with the data-defined reference through an energy-dependent attenuation baseline $\beta_0$ in central to mid-central collisions and a centrality-dependent modification of $k_\beta$ in more peripheral collisions, with only a weak dependence on $\sqrt{s_{NN}}$, providing a non-trivial and over-constrained determination of the effective scaling response. The multiplicity dependence of the scaling parameters, characterized by increasing $\zeta_{\rm rf}$ with multiplicity and a growing $\zeta_{\rm hs}$ toward peripheral collisions, reflects the interplay of EOS-driven collective expansion, finite system lifetime, and late-stage hadronic dynamics. Collectively, these results demonstrate that the scaling framework provides a quantitative, constraint-driven probe of the hydrodynamic response, enabling the disentanglement of the coupled contributions to azimuthal anisotropy.

\vspace{10pt}
\section*{Acknowledgement}
The author gratefully acknowledges Huichao Song for providing the hydrodynamic simulation results, which are based on calculations reported in Ref.~\cite{Zhao:2017yhj}.
\bibliography{pid-refs-hydro}
%
\end{document}